# A Hardware-aware Hopfield Network with a Nonlinear Memristor Array for Robust Associative Memory with Superlinear Capacity

Younghyun Lee[1], Hakseung Rhee[2], Unhyeon Kang[1,3], Seungmin Oh[1,4], Kyungmin Lee[1,5], Hyun Jae Jang[1], Seongsik Park[1], YeonJoo Jeong[1], Inho Kim[1], Jong Keuk Park[1], Kyung Min Kim[2*], Suyoun Lee[1,6*]

[1]Center for Semiconductor Technology, Korea Institute of Science and Technology (KIST), 5 Hwarang-ro 14-gil, Seongbuk-gu, Seoul, 02792, Republic of Korea

[2]Department of Materials Science and Engineering, Korea Advanced Institute of Science and Technology (KAIST), 291 Daehak-ro, Yuseong-gu, Daejeon 34141, Republic of Korea

[3]Department of Materials Science and Engineering, Seoul National University, Seoul 08826, Korea

[4]Department of Physics and Astronomy, Seoul National University, Seoul 08826, Korea

[5]Department of Electrical Engineering, Korea University, Seoul 02841, Korea

[6]Nanoscience and Technology, KIST School, University of Science and Technology, Seoul 02792, Korea

*Correspondence and requests for materials should be addressed to K.M.Kim and S.Lee.
(E-mail: K.M.Kim: km.kim@kaist.ac.kr, S.Lee: slee_eels@kist.re.kr)

Associative memory retrieves complete patterns from partial or corrupted inputs and constitutes a primitive form of generative inference. Classical Hopfield networks (CHN) provide a canonical framework for associative memory but suffer from limited memory capacity. Recently, modern Hopfield networks (MHN) were introduced to achieve higher capacity by using explicit pattern-wise storage and neurons with the softmax activation function, which makes the MHN vulnerable to noise and the hardware implementation complicated due to its network size varying with the number of stored patterns. Here, we introduce a hardware-aware Hopfield network (HHN), in which the intrinsic nonlinear current-voltage characteristics of a charge-trap memristor are leveraged to engineer the energy landscape of the HN, increasing the memory capacity. Using a 25 × 25 nonlinear memristor array, we demonstrate reliable reconstruction of corrupted patterns with memory capacity far exceeding the classical limit ($K \sim 0.14N$, where $N$ is the number of neurons). The HHN preserves Hopfield-type energy-minimization dynamics and remains robust to synaptic conductance noise. Large-scale simulations on high-dimensional image data reveal an empirical memory capacity scaling of $K \sim 0.3 \times N^{1.2}$ under a fixed synaptic budget. These results establish HHN as a scalable hardware-native architecture for low-power associative memory and generative inference.

## Introduction

The rapid proliferation of generative artificial intelligence, driven by architectures such as the Transformer, has led to an unprecedented escalation in computational energy demand, raising fundamental concerns regarding the scalability and sustainability of current digital systems.[1-3] Associative memory can be regarded as an energy-efficient alternative, as it reconstructs complete representations from partial or noisy inputs through collective dynamics, performing direct retrieval by exploiting the structure of stored representations rather than relying on large-scale sequential generation or explicit pattern-wise computation (see **Fig. 1**a).[4,5] Classical Hopfield networks (CHNs) formalize this principle as an energy-minimization process over binary neurons with symmetric interactions, but suffer from a severely limited memory capacity ($K \sim 0.14N$, where $N$ is the number of neurons).[6]

Modern Hopfield networks (MHNs) overcome this limitation by introducing higher-order nonlinear operations into neuronal computation.[7,8] Whereas CHNs, based on Hebbian-like quadratic interactions between neurons, form relatively broad and overlapping attractor basins in the energy landscape, MHNs exploit the higher-order nonlinearity of the softmax function in the neuronal activation to sharpen the energy landscape. This principle provides the basis for surpassing the memory-capacity limit of CHNs.

Despite these algorithmic advantages, MHNs remain poorly suited for efficient hardware implementation. Its core softmax operation requires evaluating interactions between the input and all stored patterns, resulting in an $N \times K$ computational structure.[7,9] On top of this, the softmax operation involves exponentiation, global summation, and logarithmic compression, which further increases computational and hardware complexity. As $K$ increases, the memory resources required for representation and the computational cost grow accordingly. This makes the model well-suited to digital processors optimized for parallel computation, such as GPUs, but unfavorable for fixed-structure in-memory hardware.

In this work, we propose a Hardware-aware Hopfield Network (HHN), a hardware-friendly associative memory architecture, that preserves the fixed architecture of a Hopfield network while replacing the higher-order neuronal nonlinearity of MHNs with hardware-native synaptic nonlinearity. Specifically, HHN employs an $N \times N$ crossbar array in which memristor devices with intrinsic nonlinear current-voltage ($I$-$V$) characteristics are placed at each crosspoint to represent the synaptic weight matrix.[10-12] We leverage this nonlinear $I$-$V$ behavior to naturally introduce the higher-order interaction terms in the Hopfield energy, sharpening the energy landscape featuring an increased number of attractors corresponding to stored patterns.

[13,14] With a slightly modified neuronal update rule, the concept of HHN can be easily implemented by using the same crossbar structure as CHN while keeping the network size regardless of the number of stored patterns. The neuronal update rule should incorporate a nonlinear synaptic transformation that, from a feature-space perspective, effectively redefines the similarity metric between network states.

During training of HHN, a fixed $N \times N$ synaptic matrix and bias are learned by imposing an approximate fixed-point condition on the stored patterns and solving the resulting constraints through linear regression.[15,16] This approach allows the synaptic weights to be determined in a single step, distinguishing it from iterative gradient-based optimization methods, such as backpropagation[17] and equilibrium propagation.[18] Through this learning procedure, the statistical structure of multiple stored patterns is represented in a compact weight matrix whose physical size remains independent of $K$, thereby enabling a fixed architecture compatible with memristor crossbar hardware.

Furthermore, HHN distributes memories across the learned weight matrix, thereby reducing sensitivity to local conductance defects and enhancing robustness to synaptic weight noise.[17] This contrasts with explicit pattern-wise storage schemes, such as softmax-based MHN,[7] which are vulnerable to noise.[19] As a result, HHN is particularly advantageous for memristor-based hardware, where device-level variability and temporal conductance fluctuations are unavoidable, as summarized in Fig. 1c.

We experimentally demonstrate associative memory operation using a 25 × 25 nonlinear memristor array and its enhanced memory capacity by successfully retrieving 26 alphabet characters, far exceeding the limit of CHN (~ 0.14 × 25 = 3.5). Furthermore, the system reliably reconstructs patterns from inputs corrupted by up to about 10%. In large-scale simulations based on a mathematical model, we observe a capacity scaling of $K = 0.3 \times N^{1.2}$ (using a cosine similarity (CosSim) threshold of > 0.97, corresponding to a 1.5% pixel error,[20] under 20% masking), highlighting both enhanced storage capacity and robust reconstruction under device conductance noise. These results establish HHN as a practical and scalable architecture for energy-efficient in-memory associative memory.

# Results

## Hardware-aware Hopfield network (HHN) system for pattern association

The central idea of HHN is to exploit the intrinsic nonlinear *I-V* characteristics of memristive synapses as the physical source of higher-order interaction terms in the Hopfield energy. Because this synapse-driven nonlinearity is fundamentally different from the linear pairwise interaction assumed in CHN, the original Hopfield energy and update rule can no longer be used directly. A new nonlinear energy function and a corresponding neuron update rule must therefore be formulated to ensure stable retrieval dynamics.

We assume that the nonlinear synaptic response function is given by $I=W\rho(V)$, where $W$ is a constant independent of $V$. In addition, let $u_i$ and $s_i$ denote the internal state variable and the output of neuron $i$, respectively, with $s_i = g(u_i) = tanh(u_i)$. Here, $g(u_i)$ is the activation function of a neuron. In HHN crossbar array, we assume a system where the continuous-time dynamics are written as $\tau \, {du_i}/{dt} = -u_i + h_i$, where $\tau$ and $h_i$ are the characteristic time constant of a neuron and the time-discrete variable representing the activation of a neuron, respectively. Here, $h_i$ is given by a product of two factors: $h_i(t_n) = u_i(t_{n-1} + \Delta t) = \rho'(s_i(t_{n-1})) \times \{\sum_j W_{ij}\rho(s_j(t_{n-1})) + b_i\}$, where $n$, $\Delta t$, and $\rho'(s)$ are a positive integer, infinitesimal time step for update of neuronal state, and $d\rho(s)/ds$, respectively.[21] In the hardware HHN, the second factor is computed naturally by the crossbar array of nonlinear memristors. Then, the update rule of the neuron can be expressed by the following equation (Eq. (1)).

$$s_i(t_n) = g\big(h_i(t_n)\big) = tanh\left[\rho'(s_i(t_{n-1})) \times \left\{\sum_j W_{ij}\rho(s_j(t_{n-1})) + b_i\right\}\right] \qquad (1)$$

Unlike the classical Hopfield network, where the weights are typically determined by a Hebbian correlation rule, the HHN weight matrix is obtained from a fixed-point regression framework. Specifically, the stored patterns are required to satisfy an approximate fixed-point condition under the nonlinear HHN dynamics, and the resulting constraints are solved by regularized regression for each neuron. In the auto-associative HHN, which adopts a recurrent architecture in which the output of each neuron is fed back as input to the network, each stored pattern serves as a regression sample. This procedure compactly encodes the statistical structure of the patterns into a fixed $N \times N$ synaptic matrix, whose physical dimensionality is independent of the number of stored patterns, thereby ensuring compatibility with static memristor crossbar hardware (See Supplementary Note 1 for detailed weight-matrix training process).

Importantly, the proposed system admits a Hopfield energy function in which the interaction term is defined in the transformed state space $\rho(s)$, rather than in the classical linear state space $s$. The nonlinear Hopfield energy is given by Eq. (2).

$$E(s) = -\frac{1}{2}\rho(s)^T W \rho(s) - b^T \rho(s) + \sum_i \int_0^{s_i} g^{-1}(v)dv \tag{2}$$

The time derivative of the energy satisfies ${dE}/{dt} = -\tau g'(u_i)\left|{du_i}/{dt}\right|^2 \leq 0$ (see Supplementary Note 2 for detailed energy convergence).[22] This result shows that the nonlinear synaptic interactions reshape the energy landscape without destroying the convergence property of the Hopfield network. In other words, HHN preserves stable energy-minimizing retrieval dynamics while introducing effective higher-order nonlinearity that sharpens the attractor basins and improves associative memory capacity.

From a representational perspective, the nonlinear synaptic response $\rho$(s) can be viewed as mapping the neuron states into a nonlinear feature space, in which pairwise associative interactions become linear in the transformed coordinates. In this sense, HHN is conceptually related to kernel methods and reservoir computing, as nonlinear state encoding enriches the effective representation before linear interaction.[13,23] However, unlike classical kernel methods, this transformation does not increase the dimensionality of the state vector; instead, it reshapes the geometry of the state space by redefining the effective similarity between patterns. As a result, the transformed interaction sharpens attractor basins and reduces overlap among stored memories, providing an intuitive explanation for the improved storage capacity of HHN compared with conventional linear Hopfield networks while preserving compatibility with static crossbar hardware (see Supplementary Note 3).

**Electrical characteristics of the nonlinear memristor array**

To implement the proposed HHN, we employ a nonlinear charge-trap memristor array as the physical synapse for associative memory operation. The device is chosen to provide the key electrical characteristics required for HHN, including analog programmability, self-rectifying behavior, and, most importantly, a strongly nonlinear *I–V* response that can be directly exploited in associative update dynamics. Figs. 2a and 2b show the optical image of the fabricated array and the cross-sectional TEM image, respectively. The device consists of a Pt/$Ta_2O_5$/$NbO_x$/$Al_2O_3$/Ti on a $SiO_2$ substrate, in which $Al_2O_3$, $NbO_x$, and $Ta_2O_5$ serve as the tunneling, charge-trap, and blocking layers, respectively.[12,24,25] More details on the device fabrication process are provided in the Methods section.

Fig. 2c shows characteristic current-voltage (*I–V*) curves measured under various compliance currents ($I_{cc}$), demonstrating tunable, nonvolatile analog conductance. The analog switching behavior is attributed to modulation of the Schottky barrier height induced by electron trapping and detrapping within the $NbO_x$ layer.[12] Fig. 2d shows the *I–V* characteristics of randomly selected cells in the array measured at $I_{cc}$ = 1 μA, indicating that the device-to-device uniformity is good enough for the application to 25 × 25 HHN array in this work. Fig. 2e demonstrates stable multilevel conductance programmability at the reference voltage of 3 V, supporting the representation of multiple analog synaptic states within the array.

For the application to HHN, a key requirement of the nonlinear memristor is that its functional dependence on the applied voltage remains invariant across different resistance states. Fig. 2f shows that the *I–V* characteristics of our device are well described by the Schottky-type conduction model across different resistance states, satisfying the above requirement (see the Supplementary Fig. 1 for fitting results). It is noteworthy that, in conventional in-memory computing, such non-Ohmic behavior is generally regarded as a non-ideality because it complicates linear vector-matrix multiplication (VMM).[26] In the present work, however, we exploit the intrinsic nonlinearity as a computational advantage. Furthermore, the device exhibits self-rectifying characteristics, which are advantageous for passive array operation by suppressing leakage current without requiring an additional selector. Together, the self-rectifying behavior, multilevel analog programmability, and nonlinear conductance characteristics make the memristor array well suited for HHN implementation. Thus, its significance here lies in enabling a fixed-footprint nonlinear synaptic array for hardware-efficient associative memory.

**Auto-associative pattern retrieval**

For experimental demonstration, we implement the above HHN formalism in a 25 × 25 crossbar array with nonlinear charge-trap memristors. To match this architecture, we use 5 × 5 pixel alphabet images corresponding to all 26 letters (A–Z; $K$ = 26) as stored patterns. Each image is reshaped into a 1 × 25 vector, so that each Hopfield neuron represents one pixel. The trained weights are then mapped and quantized into target conductance values for hardware implementation (see Supplementary Fig. 2 for the detailed weight matrix and hardware setup).

**Fig. 3**a shows the target conductance matrix corresponding to the trained HHN weights, while Fig. 3b presents the experimentally measured conductance matrix at 3 V after programming the memristor array. Note that two separate memristor arrays are used to encode positive and negative weight components, respectively. As shown in Fig. 3c, the measured and target conductance values yield a standard deviation-to-mean ratio of 5.66%, confirming that the array can represent the trained weight matrix with sufficient accuracy for associative memory operation.

We next examine the retrieval behavior of the programmed HHN. A noise-free input pattern (e.g., the letter A) is encoded into a voltage vector of +3 V and -3 V, which is applied to the two memristor arrays representing the positive and negative weight components, respectively.[27,28] Retrieval results for the full alphabet set (A-Z) are shown in Fig. 3d, demonstrating accurate recall of all stored patterns and validating the operation. The retrieval results for the full alphabet set (A–Z) are summarized in Fig. 3d, demonstrating accurate recall of all stored patterns and validating the operation of the nonlinear memristor array as a physical associative memory system.

To benchmark retrieval performance, we compare HHN with CHN and MHN using random patterns. In this analysis, MHN employs the softmax activation function.[7] As shown in Fig. 3e, CHN fails to reliably retrieve patterns once the number of stored patterns exceeds four, reflecting the limited capacity of linear associative dynamics. By contrast, MHN maintains stronger retrieval performance owing to its sharper energy landscape, albeit at the cost of explicit pattern-wise storage and computation that scales with $K$. HHN achieves comparable retrieval performance while retaining a fixed $N \times N$ hardware footprint. These results indicate that device-level nonlinearity can reshape network dynamics to enhance memory capacity without sacrificing crossbar compatibility.

We further examine robustness to synaptic perturbation by introducing conductance noise into the weight matrix. Fig. 3f shows the cosine similarity (CosSim) as a function of the normalized conductance variation ($\sigma/\mu$) at $N = 25$ and $K = 30$. As the conductance variation increases, MHN degrades markedly, whereas HHN maintains high CosSim. This difference arises from their distinct memory representations: MHN stores target patterns explicitly, making recall directly sensitive to synaptic corruption, whereas HHN distributes memory across a learned $N \times N$ weight matrix and is therefore inherently more tolerant to conductance variation (see Supplementary Fig. 3 for simulated retrieval under synaptic noise). Such robustness is particularly important for memristor hardware, where device variation, programming error, and temporal conductance noise are unavoidable.

Next, we investigate the reconstruction ability of HHN for partially corrupted patterns (see **Fig. 4**a). A 5 × 5 alphabet pattern with a randomly positioned 2 × 2 masked region is iteratively updated, and its evolution is tracked over iterations. Representative examples (H, O, and P) are shown in Fig. 4b, where intermediate current maps at iterations 1, 5, and 10 illustrate the retrieval process. While early iterations retain features of the corrupted input, the current distribution progressively converges towards the stored pattern, reflecting suppression of inconsistent pixel states and reinforcement of the learned representation. These results demonstrate that the HHN enables iterative associative recovery directly in hardware, highlighting its potential for hardware-native generative inference. Fig. 4c quantifies this retrieval process for $N = 25$ and $K = 26$ with a mask ratio of 0.16 (= 4/25) and conductance variation of 5.66%. As iterations proceed, the CosSim increases monotonically, while the Hopfield energy (Eq. (2)) converges to a steady value, confirming that the programmed HHN follows the expected energy-minimization dynamics during incomplete-pattern recovery.

**Scalability of HHN for large-scale pattern association**

To assess the scalability of HHN beyond the experimentally demonstrated 25 × 25 array, we perform large-scale simulations on both the Fashion-MNIST dataset (f-MNIST) and random pixel patterns, incorporating device-level conductance variation uniformly distributed within ± 5.66%, as observed experimentally. **Fig. 5**a shows a representative pattern evolution at corruption levels of 0.2, 0.3, and 0.4, where initially degraded inputs are progressively refined, eventually converging toward the target images through iterative updates. As shown in Fig. 5b, the CosSim increases monotonically and approaches unity, with higher corruption levels requiring more iterations. These results confirm the robust reconstruction capability of HHN in large-scale settings.

We next use random pixel patterns (Fig. 5c) to estimate the intrinsic storage capacity of the network. The resulting capacity scaling is summarized in Fig. 5d using a CosSim threshold of > 0.97 (1.5% pixel error[20]). For a fair comparison across CHN, MHN, and HHN, the storage capacity is evaluated under an equal synaptic budget, $S = N^2$, corresponding to the same total number of synaptic elements in an $N \times N$ matrix. Under this constraint, HHN shows a clear advantage over the other networks. Fitting the HHN results yields an empirical scaling of approximately $0.3 \times N^{1.2}$, indicating superlinear growth of storage capacity with neuron number (see Supplementary Fig. 4 for details of capacity extraction).[20] By contrast, although MHN can provide high algorithmic capacity, its memory representation and computation are inherently $K$-dependent, making the fixed-synaptic-budget comparison particularly relevant from a hardware perspective.

Furthermore, to assess the scalability of reconstruction speed, we investigate the number of iterations required to reconstruct the target image with varying $K$ (= 200, 500, and 1000) at a fixed $N = 900$. In Fig. 5e, note that CosSim converges very fast within ten iterations irrespective of $K$. In addition, it shows perfect pattern reconstruction for $K = 200$ and 500, although CosSim for $K = 1000$ saturates at a rather low value of ~ 0.92 due to the capacity limit of the network. In the Supplementary Fig. 5, we present additional retrieval results under different masking states -1, 0, and 1, confirming consistent trends across conditions. Notably, all simulations include 5.66% synaptic weight noise, demonstrating that the scalability of HHN is preserved under hardware-relevant perturbations. Taken together, these results demonstrate that HHN enables large-scale associative retrieval under a fixed synaptic footprint while providing enhanced storage capacity and robustness to synaptic noise.

## Conclusion

In this work, we introduce a hardware-aware Hopfield network that bridges the gap between the high capacity of nonlinear associative memory models and the efficiency of fixed-footprint in-memory computing architectures. A central insight is that device nonlinearity, often regarded as a non-ideality in analogue crossbar systems, can instead be harnessed as a computational resource. In the HHN, the intrinsic nonlinear response of the memristor effectively introduces the higher-order interaction terms to the Hopfield energy, sharpening attractor basins and reducing interference among stored patterns. This enables enhanced capacity without the $N \times K$ scaling characteristic of MHNs, making this approach well-suited to practical in-memory computing hardware. By embedding effective nonlinearity into a memristor crossbar platform, HHN preserves an $N \times N$ synaptic structure while enabling associative retrieval beyond the capability of the classical Hopfield network, which relies on purely linear coupling elements. Experimentally, we demonstrate pattern reconstruction using a 25 × 25 nonlinear memristor array, confirming operation as a physical associative memory rather than a passive analogue storage system. Large-scale simulations further highlight the strong scalability of HHN, with an empirical storage-capacity scaling of approximately $0.3 \times N^{1.2}$ under a fixed synaptic footprint.

Another key advantage of HHN is its robustness to synaptic perturbation. As associative memory is distributed across a learned $N \times N$ weight matrix rather than stored explicitly as individual target patterns, the network is inherently less sensitive to local conductance variation. This property is particularly advantageous for memristor-based hardware, where variability, programming error, and temporal noise are unavoidable. Together, these results suggest that HHN provides a practical route toward low-power, hardware-native associative and generative computation.

A remaining limitation of the present work is that the nonlinear synaptic function $\rho$ is not yet independently tunable at the device level. In the present implementation, $\rho$ is obtained by fitting the measured memristor response, such that the exponential sharpening parameter is determined by the intrinsic device characteristics rather than optimized directly for associative memory performance. An important direction for future work is therefore to identify the optimal nonlinear parameter for memory capacity, robustness, and convergence, and to realize this target response through device optimization or architectural modification.[29] Such efforts would enable tighter co-design between device nonlinearity and network dynamics, and may ultimately extend the present framework toward a more general device-aware learning rule in which the synaptic update model is explicitly matched to the programmable nonlinear response of the hardware.

Overall, HHN establishes that nonlinear memristor arrays can support not only compact in-memory storage, but also stable and scalable associative computation. More broadly, this work suggests that the physical properties of emerging devices can be incorporated directly into neural-network design, rather than treated merely as constraints to be compensated for. By leveraging device nonlinearity as a functional component of the network, HHN provides a practical framework for low-power associative memory and a promising step toward hardware-native generative computing.

## Methods

### Device fabrication

For a nonlinear memristor array, the integrated device consisted of Pt/$Ta_2O_5$/$NbO_x$/$Al_2O_3$/Ti on a $SiO_2$ substrate. The array was integrated by the following steps: First, a Ti bottom electrode was formed on a $SiO_2$ substrate by electron-beam evaporation (KVE-E2000), followed by a lift-off process. Next, an $Al_2O_3$ layer was deposited by thermal atomic layer deposition (ALD) at 180°C using trimethylaluminum and $O_3$ as the Al precursor and oxygen source, respectively. The $NbO_x$ layer was deposited by reactive sputtering (Daeki Hitech co-sputtering system) at 170°C in an Ar and $O_2$ mixed gas ambient environment using an Nb target. Next, the $Ta_2O_5$ layer was deposited using plasma-enhanced ALD at 220°C using Tris(diethylamido)(tert-butylimido)tantalum(V) and $O_2$ plasma for the Ta precursor and the oxidant, respectively.

Finally, a 50-nm-thick Pt top electrode was deposited by e-beam evaporation and patterned by a lift-off process. The line width for the active area in the array was 5 μm.

**Electrical measurements**

Electrical characterization was performed using a Keithley 4200A-SCS and an ArcOne f-MDPE controller. The pulse measurements were obtained by a Keithley 4200A-SCS and an ArcOne. During the measurement, the TE was biased, and the BE was grounded.

## Data availability

All the data supporting the findings of this study are available within the article and its Supplementary Information.

## Code availability

The simulation results were processed using Python software. All the relevant codes are available from the corresponding author upon reasonable request.

## Acknowledgments

This work was supported by the Korea Institute of Science and Technology (KIST, Grant No. 26E0020), National Research Foundation of Korea (NRF, Grant No. RS-2025-24533987), and the National Research Council of Science & Technology (NST, Grant No. GTL24041-000).

## Author contributions

S.L. and Y.L. conceived the hardware-aware Hopfield network. Y.L., H.J.J. and S.P. developed the learning rule for the weight matrix and the neuron update dynamics. Y.J., I.K., J.K.P. and K.M.K. contributed to the design of the nonlinear memristor device. Y.L. and H.R. fabricated the memristor array, and Y.L., H.R. and U.K. carried out device characterization. Y.L., S.O. and K.L. performed large-scale HHN simulations. All authors discussed the results and contributed to writing the manuscript.

## References

1 Vaswani, A. *et al.* Attention is all you need. *Advances in neural information processing systems* **30** (2017).

2 Patterson, D. *et al.* Carbon emissions and large neural network training. *arXiv preprint arXiv:2104.10350* (2021).

3 Brown, T. *et al.* Language models are few-shot learners. *Advances in neural information processing systems* **33**, 1877-1901 (2020).

4 Marr, D. Simple memory: a theory for archicortex. *Philosophical Transactions of the Royal Society of London. B, Biological Sciences* **262**, 23-81 (1971).

5 McClelland, J. L., McNaughton, B. L. & O'Reilly, R. C. Why there are complementary learning systems in the hippocampus and neocortex: insights from the successes and failures of connectionist models of learning and memory. *Psychological review* **102**, 419 (1995).

6 Hopfield, J. J. Neural networks and physical systems with emergent collective computational abilities. *Proceedings of the National Academy of Sciences* **79**, 2554-2558 (1982).

7 Ramsauer, H. *et al.* Hopfield networks is all you need. *ArXiv* (2020).

8 Krotov, D. & Hopfield, J. J. Dense associative memory for pattern recognition. *Advances in Neural Information Processing Systems* **29** (2016).

9 Alonso, N. & Krichmar, J. L. A sparse quantized hopfield network for online-continual memory. *Nature Communications* **15**, 3722 (2024).

10 Jeon, K. *et al.* Purely self-rectifying memristor-based passive crossbar array for artificial neural network accelerators. *Nature Communications* **15**, 129 (2024).

11 Ren, S. G. *et al.* Self-rectifying memristors for three-dimensional in-memory computing. *Advanced Materials* **36**, 2307218 (2024).

12 Kim, G. *et al.* Retention secured nonlinear and self-rectifying analog charge trap memristor for energy-efficient neuromorphic hardware. *Advanced Science* **10**, 2205654 (2023).

13 Jaeger, H. & Haas, H. Harnessing nonlinearity: Predicting chaotic systems and saving energy in wireless communication. *Science* **304**, 78-80 (2004).

14 Appeltant, L. *et al.* Information processing using a single dynamical node as complex system. *Nature Communications* **2**, 468 (2011).

15 Tapson, J. & van Schaik, A. Learning the pseudoinverse solution to network weights. *Neural Computation* **45**, 94-100 (2013).

16 Zhu, V. & Rosenbaum, R. Learning fixed points of recurrent neural networks by reparameterizing the network model. *Neural Computation* **36**, 1568-1600 (2024).

17 He, C. *et al.* A hardware-adaptive learning algorithm for superlinear-capacity associative memory on memristor crossbars. *Nature Communications* (2026).

18 Yi, S.-i., Kendall, J. D., Williams, R. S. & Kumar, S. Activity-difference training of deep neural networks using memristor crossbars. *Nature Electronics* **6**, 45-51 (2023).

19 Agliari, E. & De Marzo, G. Tolerance versus synaptic noise in dense associative memories. *The European Physical Journal Plus* **135**, 1-22 (2020).

20 Amit, D. J., Gutfreund, H. & Sompolinsky, H. Storing infinite numbers of patterns in a spin-glass model of neural networks. *Physical Review Letters* **55**, 1530 (1985).

21 Krotov, D. & Hopfield, J. Large associative memory problem in neurobiology and machine learning. *arXiv preprint arXiv:2008.06996* (2020).

22 Hopfield, J. J. Neurons with graded response have collective computational properties like those of two-state neurons. *Proceedings of the National Academy of Sciences* **81**, 3088-3092 (1984).

23 Oh, S. *et al.* Optimizing reservoir connectivity: A path to high-performance liquid state machines. *Neurocomputing*, 132037 (2025).

24 Lee, Y. *et al.* Flexible self-rectifying synapse array for energy-efficient edge multiplication in electrocardiogram diagnosis. *Nature Communications* **16**, 4312 (2025).

25 Kim, G. *et al.* Band and Field Coengineered Charge Trap Memristor via Au Nanoparticle Layer for Programming Speed Enhancement. *Small Science* **5**, e202500309 (2025).

26 Joksas, D. *et al.* Nonideality-aware training for accurate and robust low-power memristive neural networks. *Advanced Science* **9**, 2105784 (2022).

27 Aguirre, F. *et al.* Hardware implementation of memristor-based artificial neural networks. *Nature Communications* **15**, 1974 (2024).

28 Yao, P. *et al.* Fully hardware-implemented memristor convolutional neural network. *Nature* **577**, 641-646 (2020).

29 Chiu, F.-C. A review on conduction mechanisms in dielectric films. *Advances in Materials Science and Engineering* **2014**, 578168 (2014).

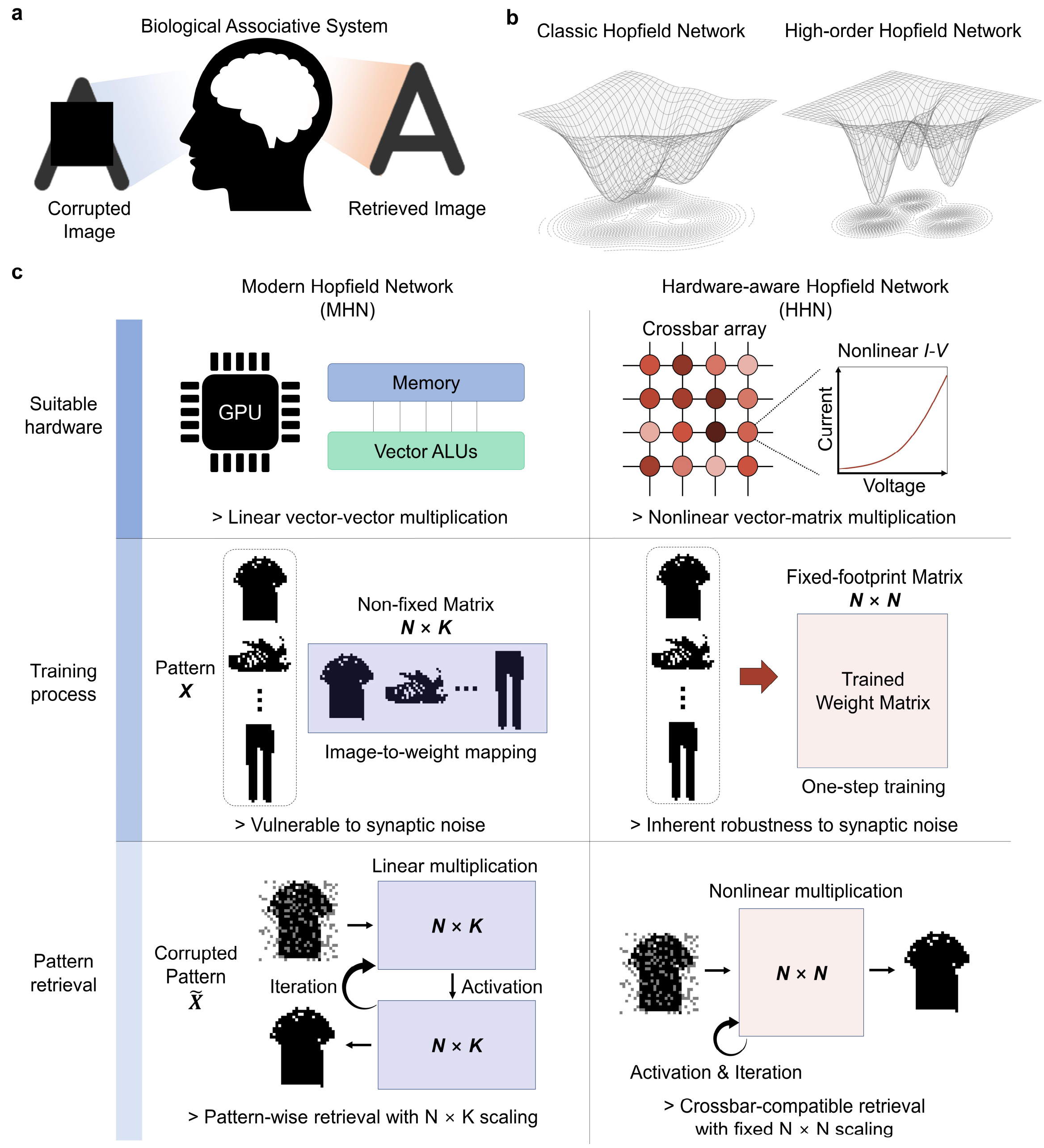

**Fig. 1 Hardware-aware associative memory architecture. a,** Biological associative memory for retrieving complete images from corrupted inputs, **b,** Energy landscape comparison between the classical Hopfield network and a higher-order Hopfield network. A sharper landscape enables a larger associative memory capacity. **c,** Comparison between the Modern Hopfield Network (MHN) and the proposed Hardware-aware Hopfield Network (HHN), highlighting differences in the source of nonlinearity, storage/update mechanism, hardware suitability, and robustness.

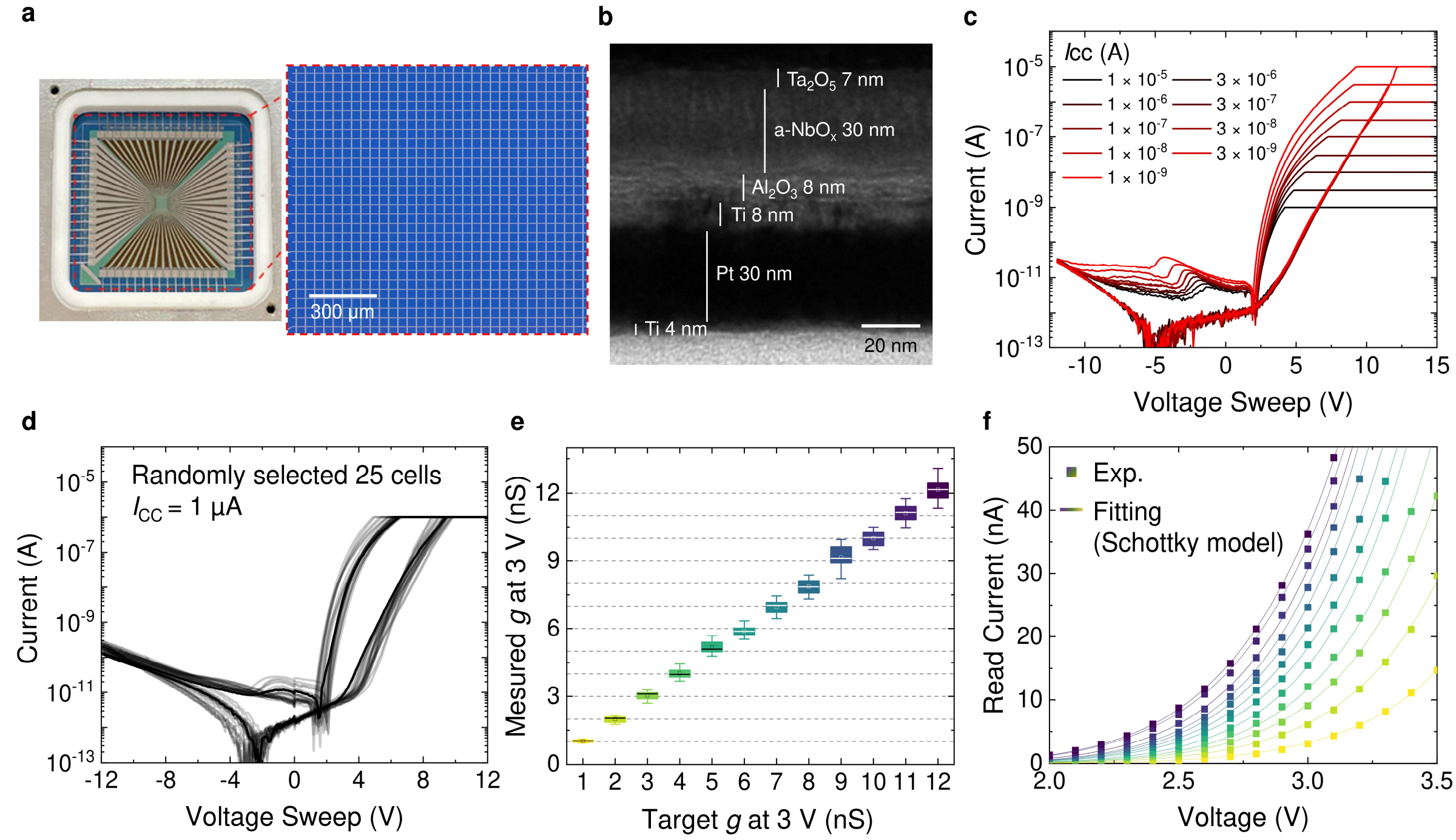

**Fig. 2 Electrical characteristics of the nonlinear memristor array. a,** Optical images of the nonlinear memristor array. **b,** Cross-sectional transmission electron microscopy (TEM) image of the device. **c,** *I–V* curves of the device with various compliance currents ($I_{CC}$). **d,** *I–V* curves of randomly selected 25 cells in the array under 1 μA of $I_{CC}$. **e,** 12 levels of multilevel conductance states at the reference voltage, 3 V, with 15 cells. **f,** *I–V* curves at the read voltage region at each analog state with Schottky fitting results.

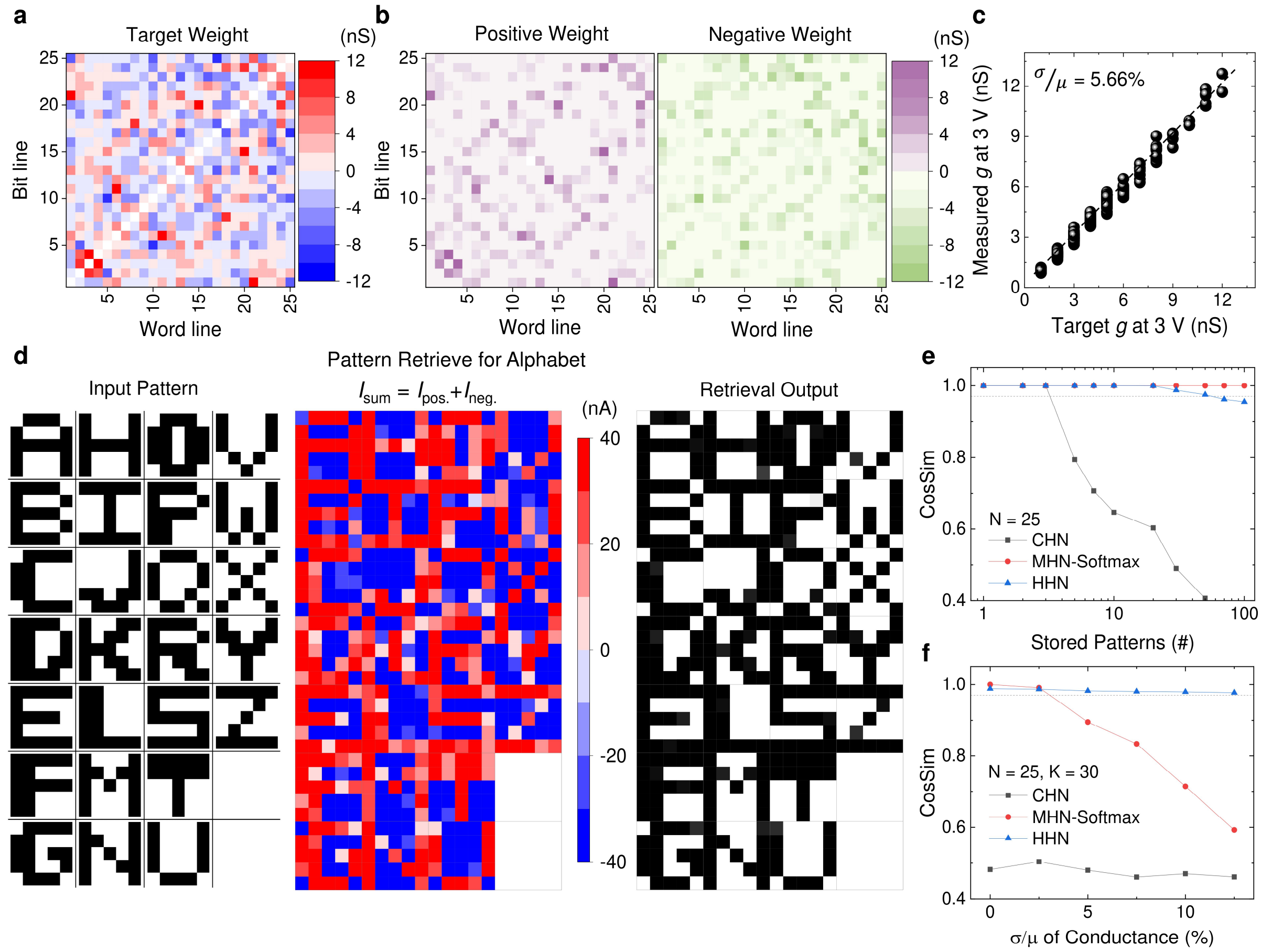

**Fig. 3 Hardware-aware Hopfield network (HHN) system for pattern association. a,** Target conductance matrix (25×25) representing the trained weights for stored alphabet patterns (5×5 pixel images). **b,** Experimentally measured conductance matrix at 3 V after programming the target weight matrix. **c**, Distribution of errors between the target and measured current at each matrix element. **d,** Retrieval process for a given input pattern and the corresponding summed-current output ($I_{sum}$=$I_{pos.}$+$I_{neg.}$), and the output image after activation, demonstrating successful pattern retrieval for alphabet patterns. **e,** Cosine similarity (CosSim) comparison among the Classical Hopfield Network (CHN), SoftMax-based Modern Hopfield Network (MHN-SoftMax), and Hardware-aware Hopfield Network (HHN) as a function of the number of stored patterns for $N$ = 25. **f,** CosSim comparison under synaptic conductance noise, showing the robustness of HHN.

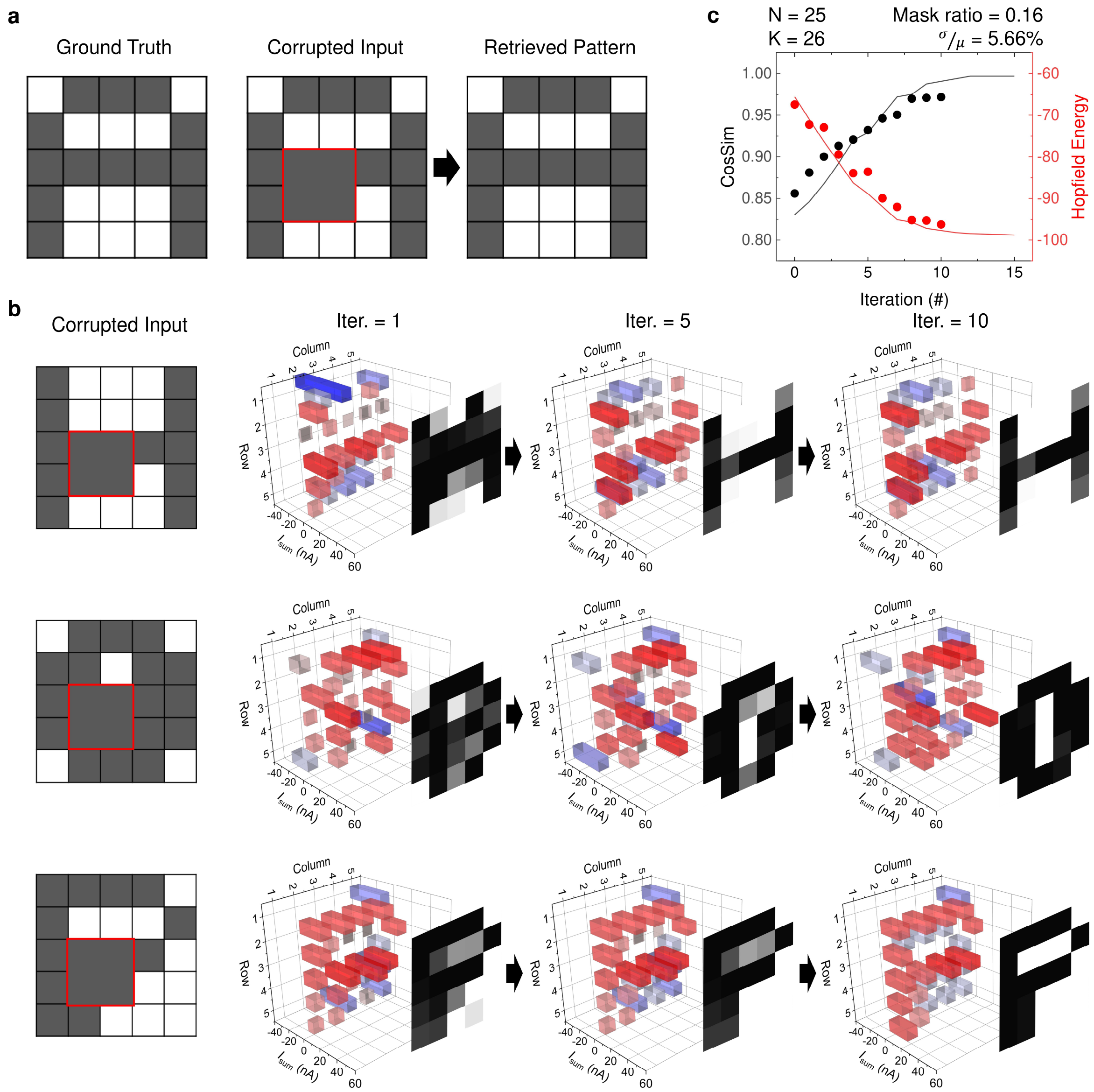

**Fig. 4 Experimental demonstration of incomplete-pattern recovery with the HHN system.** **a,** Representative example of pattern recovery by HHN from a partially masked input pattern. Ground-truth, corrupted input, and retrieved pattern are shown. **b,** Iteration-dependent evolution of CosSim and Hopfield energy for the retrieval process. **c,** Iterative recovery process for multiple corrupted input patterns (H, O, and P). Retrieved current maps at different iteration steps show progressive convergence toward the stored patterns.

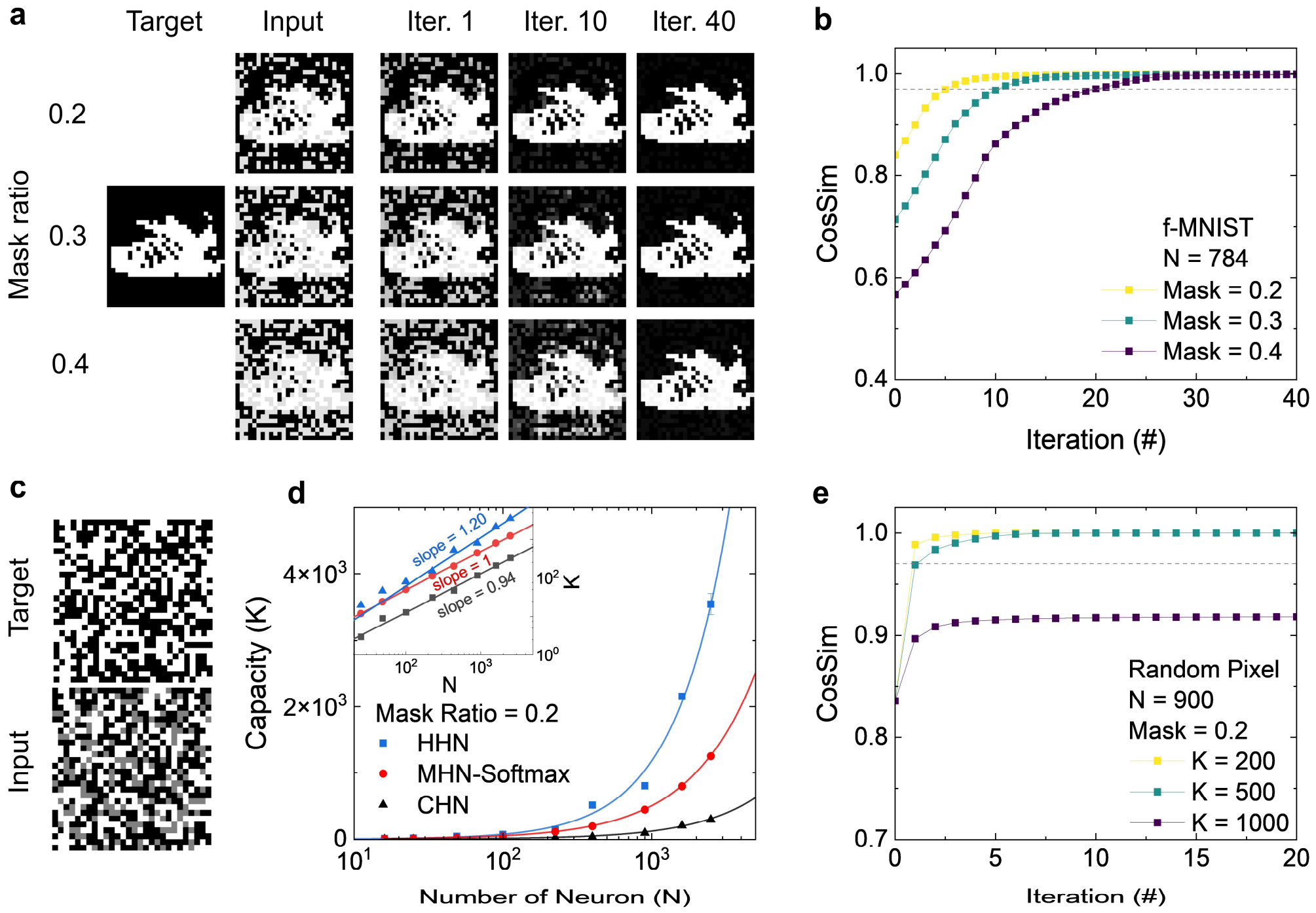

**Fig. 5 Scalability of HHN for large-scale pattern association. a,** Representative recovery evolution for incomplete f-MNIST patterns under mask ratios of 0.2, 0.3, and 0.4. **b,** CosSim as a function of iteration for f-MNIST retrieval under mask ratios of 0.2, 0.3, and 0.4. **c,** Random pixel patterns used for the estimation of memory capacity. **d,** Memory capacity as a function of neuron number for CHN, MHN, and HHN under an equal synaptic budget ($S = N^2$). The inset shows the curves in the log-log scale, revealing the behavior following the power law with the power of 0.94, 1, and 1.2 for CHN, MHN, and HHN, respectively. **e,** CosSim as a function of iteration for random pixel patterns with $K$ = 200, 500, and 1000 at $N$ = 900. All simulations were performed with 5.66% synaptic weight noise

# Supplementary Information

# A Hardware-aware Hopfield Network with a Nonlinear Memristor Array for Robust Associative Memory with Superlinear Capacity

Younghyun Lee[1], Hakseung Rhee[2], Unhyeon Kang[1,3], Seungmin Oh[1,4], Kyungmin Lee[1,5], Hyun Jae Jang[1], Seongsik Park[1], YeonJoo Jeong[1], Inho Kim[1], Jong Keuk Park[1], Kyung Min Kim[2*], Suyoun Lee[1,6*]

[1]Center for Semiconductor Technology, Korea Institute of Science and Technology (KIST), 5 Hwarang-ro 14-gil, Seongbuk-gu, Seoul, 02792, Republic of Korea

[2]Department of Materials Science and Engineering, Korea Advanced Institute of Science and Technology (KAIST), 291 Daehak-ro, Yuseong-gu, Daejeon 34141, Republic of Korea

[3]Department of Materials Science and Engineering, Seoul National University, Seoul 08826, Korea

[4]Department of Physics and Astronomy, Seoul National University, Seoul 08826, Korea

[5]Department of Electrical Engineering, Korea University, Seoul 02841, Korea

[6]Nanoscience and Technology, KIST School, University of Science and Technology, Seoul 02792, Korea

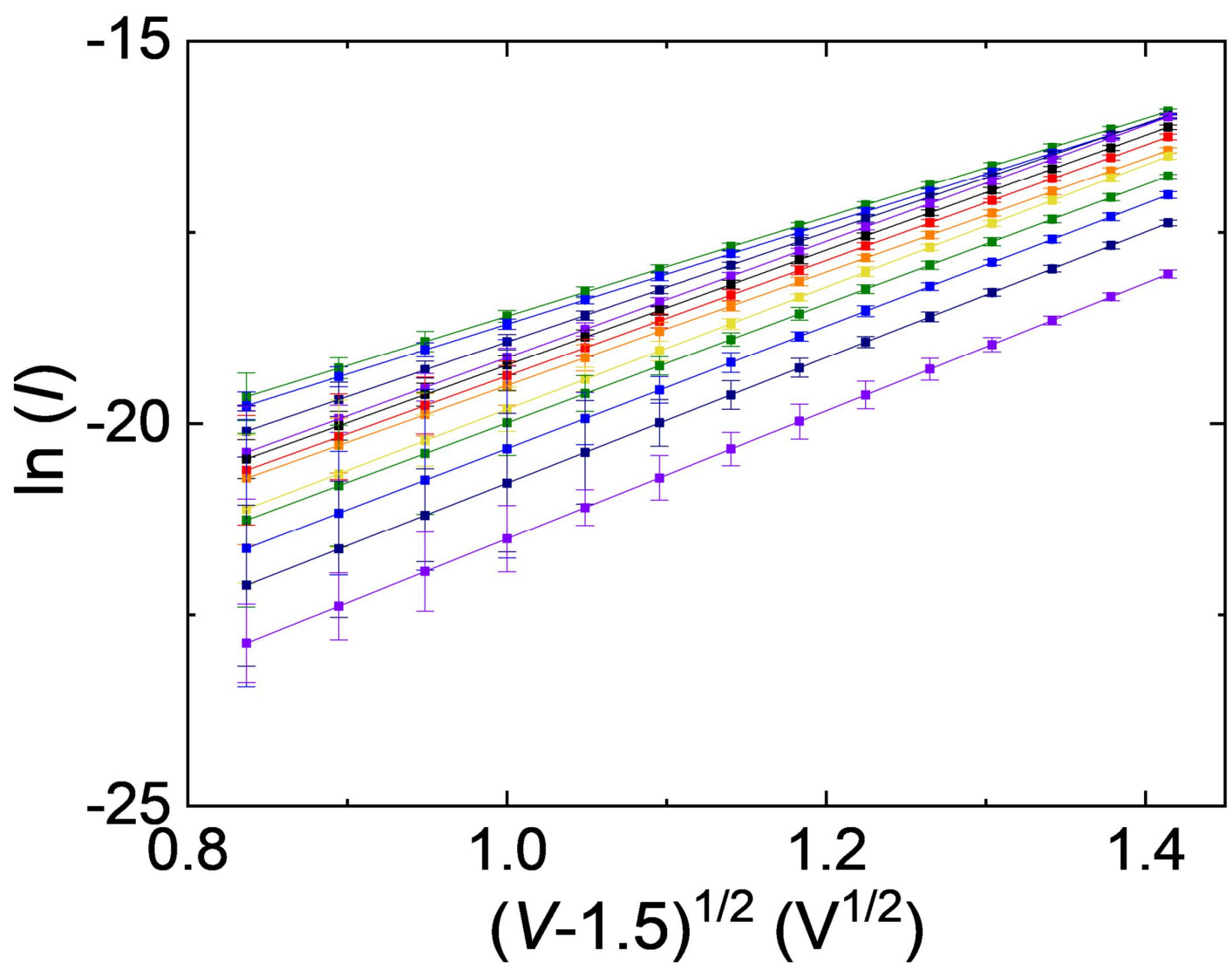

**Supplementary Fig. 1** Ln($I$) vs. $(V\text{-}1.5)^{1/2}$ plots for memristor devices (20 devices) used in the HHN in this work. The observed consistent linear behavior, $\ln(I) = \ln(I_0) + k \times (V\text{-}1.5)^{1/2}$, implies that the transport is described by the Schottky conduction regardless of the state of the memristor, which is a key requirement of a nonlinear memristor in HHN. $k$ is estimated to be $7.55 \pm 0.56$.

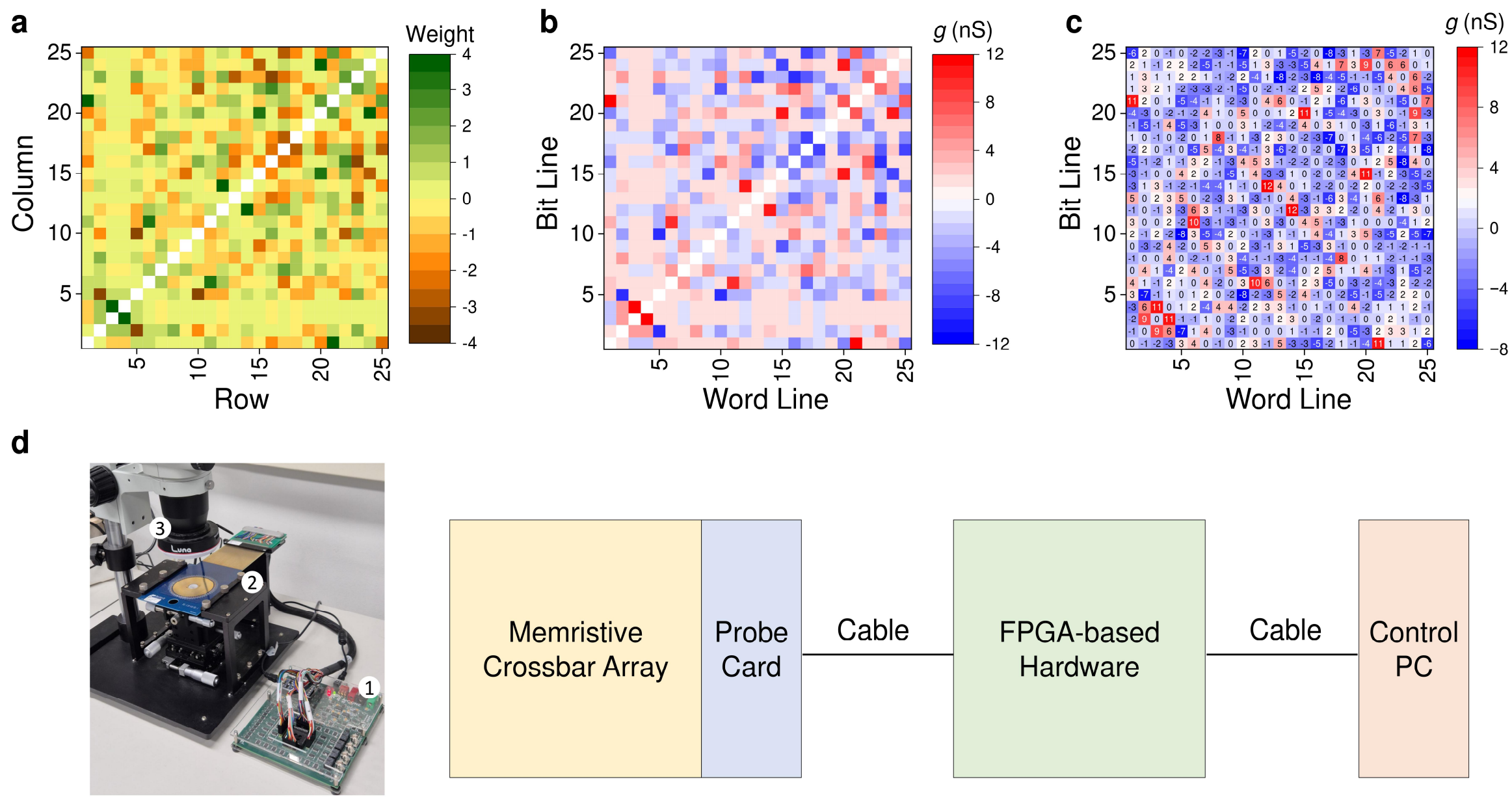

**Supplementary Fig. 2** Weight matrix construction and conductance mapping for the alphabet retrieval test. **a,** Weight matrix learned for the 26 alphabet patterns (A-Z) using the fixed-point regression method described in Supplementary Note 1. The minimum and maximum weight values are -3.33 and 3.88, respectively. **b,** Conductance matrix obtained by linearly scaling the learned weight matrix in (**a**) by a factor of $3 \times 10^{-9}$ to match the device conductance range. **c,** Quantized conductance matrix used for hardware implementation. Signed integer conductance values are obtained from the subtraction of two 25×25 memristor crossbar arrays. **d,** Measurement setup. ① is FPGA-based hardware (ArcOne, ARC Instruments Inc.), which is connected to the probe card (②), where a memristor array is loaded on the positioner using an optical microscope (③). The FPGA-based hardware is controlled by a PC.

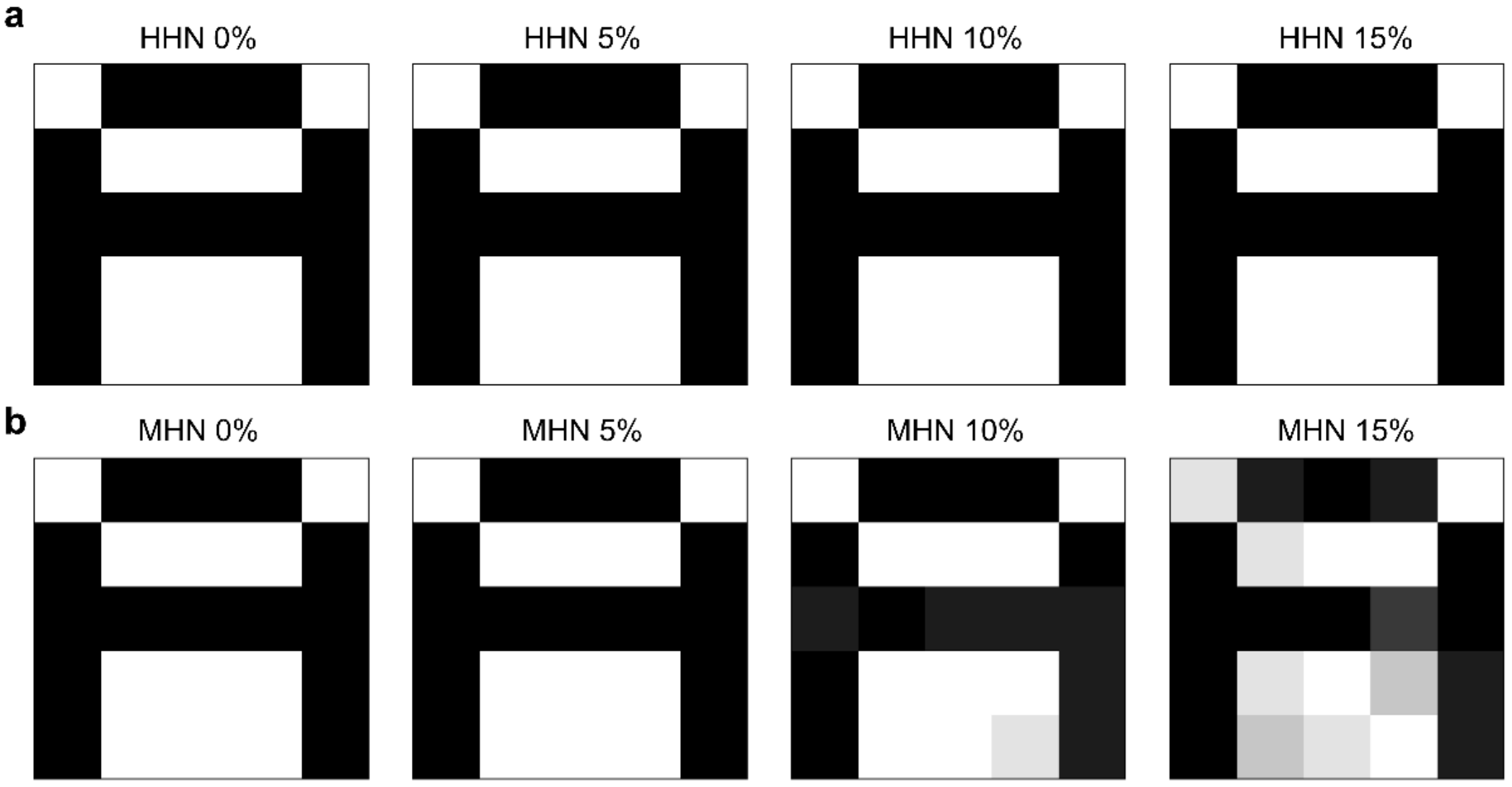

**Supplementary Fig. 3** Retrieved patterns under synaptic noise. **a,** Retrieved output of the letter "A" in HHN under conductance variations of 0%, 5%, 10%, and 15%. **b,** Retrieved output of the letter "A" in MHN under conductance variations of 0%, 5%, 10%, and 15%.

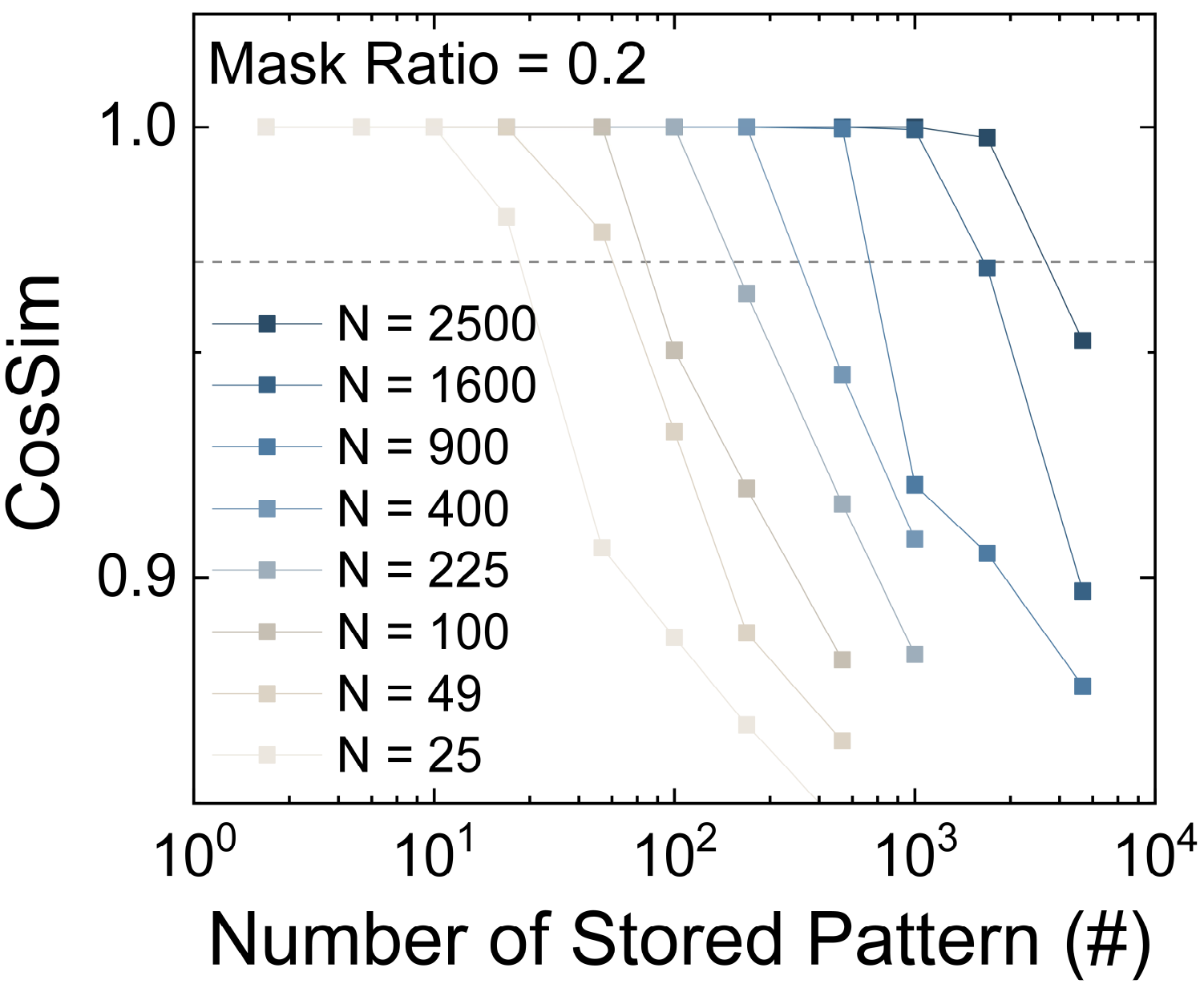

**Supplementary Fig. 4** Extraction of memory capacity using randomly generated pixel patterns. Cosine similarity as a function of the number of stored patterns for different network size $N$ under a mask ratio of 0.2. The memory capacity for each $N$ was extracted at the threshold of cosine similarity > 0.97, corresponding to a 1.5% pixel-error tolerance.

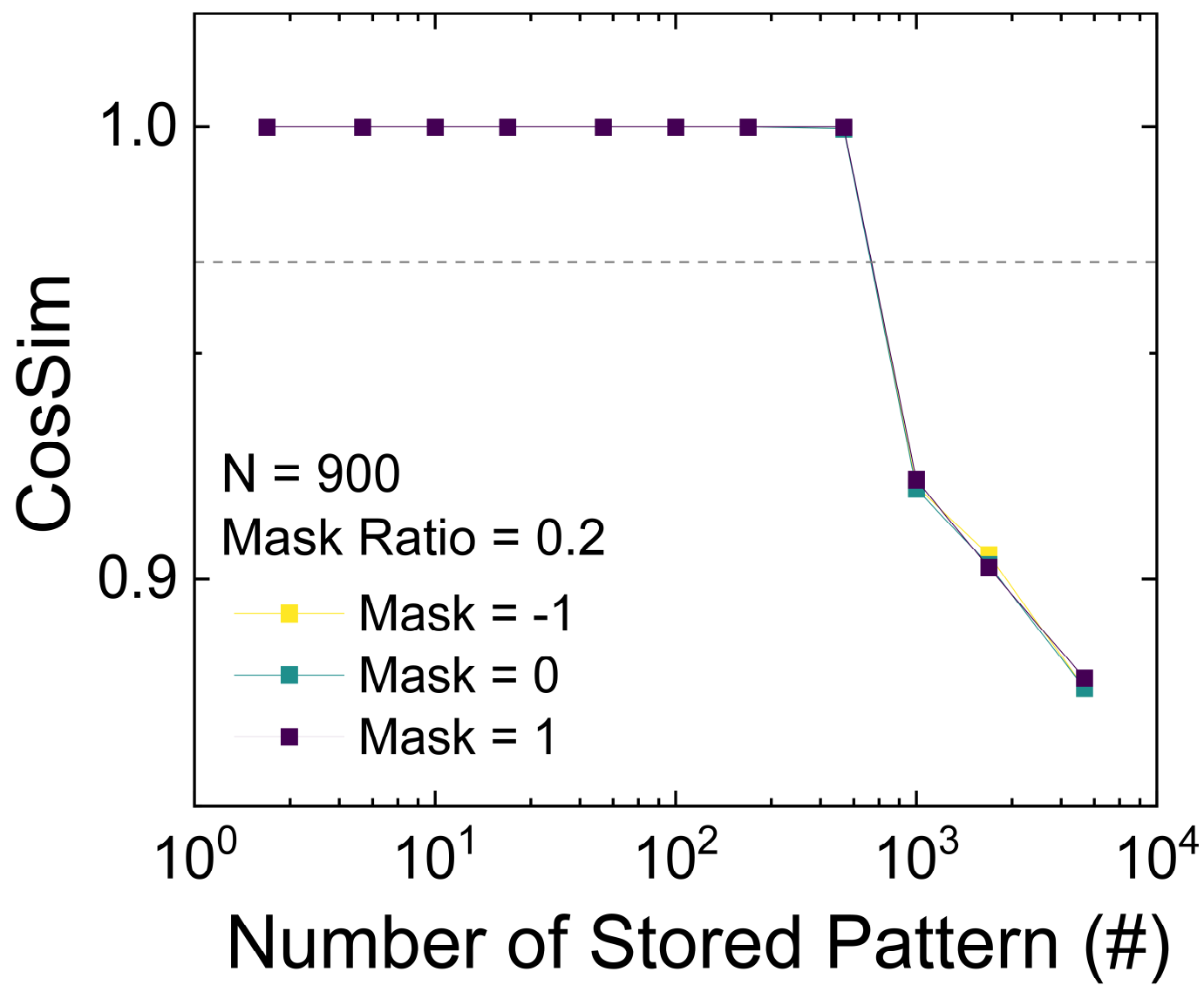

**Supplementary Fig. 5** Retrieval performance under different masking states -1, 0, and 1. Cosine similarity as a function of the number of stored patterns for $N = 900$ and a mask ratio of 0.2. The three masking states produce nearly identical retrieval trends, indicating that the storage-capacity behavior is not sensitive to the choice of masking value.

**Supplementary Note 1**

Fixed-point formulation and regression-based weight construction

To construct the weight matrix of the proposed hardware-aware Hopfield network (HHN), we use a regression-based learning rule that explicitly enforces the stored patterns to become approximate fixed points of the network dynamics.

Let the set of stored bipolar patterns be denoted by

$$t_i^{\mu} \in \{-1, +1\}, \qquad i = 1, 2, \dots, N, \text{and } \mu = 1, 2, \dots, K$$

, where $K$ is the number of stored patterns, and $N$ is the number of neurons. These patterns are obtained by transforming binary images into bipolar representations.

In HHN, the retrieval dynamics are governed by nonlinear synaptic interactions and bounded neuronal activations. Because the effective synaptic input depends on a nonlinear transformation ($\rho$) and the neuronal output is given by a bounded activation function ($g$), the exact fixed-point condition is generally nonlinear and cannot be solved analytically in closed form.

To obtain a tractable learning rule, we used the fixed-point condition around a representative operating point. Under this approximation, the fixed-point condition for neuron $i$ and stored pattern μ can be written as the following equation (1).

$$t_i^{\mu} \approx \rho'(t_i^{\mu}) \left\{ \sum_{j \neq i} W_{ij} \rho(t_j^{\mu}) + b_i \right\} \tag{1}$$

The $W_{ij}$ denotes the synaptic weight from neuron $j$ to neuron $i$, the $b_i$ is the bias term of neuron $i$, and the $\rho$ is the nonlinear synaptic function. This approximation transforms the nonlinear fixed-point condition into a linear relation with respect to the unknown parameters $W_{ij}$ and $b_i$.

For each neuron $i$, the fixed-point constraints are collected across all stored patterns. We define the target vector as follows (Eq. (2)).

$$y_i = \begin{bmatrix} t_i^1 \\ \vdots \\ t_i^K \end{bmatrix} \tag{2}$$

and the design input matrix as follows, equation (3).

$$x_i = \alpha \begin{bmatrix} \rho(t_1^1) \cdots \rho(t_{i-1}^1)\ \rho(t_{i+1}^1) \cdots \rho(t_N^1)\ 1 \\ \vdots \quad \cdots \quad \vdots \quad \vdots \quad \cdots \quad \vdots \quad \vdots \\ \rho(t_1^K) \cdots \rho(t_{i-1}^K)\ \rho(t_{i+1}^K) \cdots \rho(t_N^K)\ 1 \end{bmatrix} \tag{3}$$

, where the $i$-th neuron itself is excluded from the synaptic terms, and the last column corresponds to the bias contribution, and $\alpha = \rho'(1) = \rho'(-1)$. The unknown weight parameter vector for neuron $i$ is defined as follows (Eq. (4)).

$$w_i = \begin{bmatrix} W_{i1} \\ \vdots \\ W_{i,i-1} \\ W_{i,i+1} \\ \vdots \\ W_{iN} \\ b_i \end{bmatrix} \tag{4}$$

So that the fixed-point constraints can be written compactly by the following equation (5).

$$y_i \approx x_i w_i \tag{5}$$

Thus, for each neuron $i$, the problem reduces to a linear regression task for determining the incoming synaptic weights and the bias.

To ensure numerical stability and reduce overfitting, we solve the parameter-estimation problem using regression:

$$w_i = \operatorname{argmin}|x_i w_i - y_i| \tag{6}$$

, where the closed-form solution[1] is

$$w_i = (x_i^T x_i)^{-1} x_i^T y_i. \tag{7}$$

In practice, equation (6) is converted into an equivalent augmented least-squares problem and solved using a standard least-squares solver. The augmented system was solved using torch.linalg.lstsq, such a standard least-squares solver typically relies on numerically stable algorithms.

After solving for $w_i$, the synaptic weights and bias of neuron $i$ are extracted as follows, equation (8).

$$\begin{gathered} W_{ij} = (w_i)_j, where\ j \neq i \\ b_i = last\ element\ of\ w_i \\ W_{ii} = 0 \end{gathered} \tag{8}$$

This procedure is repeated independently for all neurons $i$ = 1, ..., N, yielding the full weight matrix $W$ and bias vector $b$.

Because each row of the weight matrix is learned independently, the resulting matrix is not guaranteed to be symmetric. However, symmetry is required for the existence of a Hopfield-type energy function. To enforce this property, we symmetrize the learned matrix as follows, equation (9).

$$W_{final} = \frac{W + W^T}{2} \tag{9}$$

**Supplementary Note 2**

Hopfield Energy and Convergence of the Nonlinear Synaptic Hopfield Network

To analyze the stability of the proposed hardware-aware Hopfield network, we consider a continuous-time dynamical system in which nonlinearities are introduced both at the neuronal and synaptic levels.

Let $u_i$ denote the internal state of neuron $i$, and let $s_i$ denote its output. The neuronal output is defined by an activation function as follows (Eq. (10)).

$$s_i = g(u_i) \tag{10}$$

In this work, we adopt

$$s_i = tanh(u_i) \tag{11}$$

, which ensures that the neuron states remain within the interval (-1, 1).

Unlike the classical Hopfield network, where synaptic interactions are linear in the neuronal outputs, we introduce a nonlinear synaptic function $\rho$. For the charge-trap memristor considered in this work, the nonlinear transformation is modeled as $\rho(s) = \text{sign}(s)\exp(k\sqrt{|s|})$, where the parameter $k$ is extracted from Schottky-conduction fitting results in Supplementary Figure 1. Here, $|s|$ is used in the square root term because the neuron state $s$ ranges from -1 to 1, while the square root term requires positive arguments. The prefactor sign($s$) ensures that $\rho(s)$ remains an odd-symmetric function, so that positive and negative states are treated with symmetric nonlinear attribution.

The synaptic output to neuron $i$ is formed as follows (Eq. (12)).

$$o_i = \sum_j W_{ij}\,\rho(s_j) + b_i \tag{12}$$

Eq. (12) represents the weighted interaction in the nonlinear transformed space, where $W_{ij}$ is a symmetric weight matrix and $b_i$ is a bias term. This term is then multiplied by $\rho'(s_i)$, introducing an additional modulation. Accordingly, the effective synaptic output can be written as the following equation (13).

$$h_i = \rho'(s_i)\left\{\sum_j W_{ij}\,\rho(s_j) + b_i\right\} \tag{13}$$

The continuous-time dynamics of the network are defined as follows, equation (14).

$$u_i + \tau \frac{du_i}{dt} = \rho'(s_i) \left\{ \sum_j W_{ij}\, \rho(s_j) + b_i \right\} \tag{14}$$

Under this definition, the updated internal state of the neuron can be written as $u_i^{t+1} = h_i^t$ and the corresponding output state is given by the following equation (15).

$$s_i^{t+1} = tanh \left[ \rho'(s_i^t) \left\{ \sum_j W_{ij}\, \rho(s_j^t) + b_i \right\} \right] \tag{15}$$

Under this neuron update rule, the Hopfield energy defined below naturally decreases during the transient dynamics. We define the following energy function (16).

$$E(s) = -\frac{1}{2}\rho(s)^T W \rho(s) - b^T \rho(s) + \sum_i \int_0^{s_i} g^{-1}(v)dv \tag{16}$$

This energy consists of three terms: the interaction energy in the transformed feature space $\rho$(s), bias contribution $b$, and a single-neuron potential term that depends on both $\rho$ and $g$.

We next compute the derivative of the energy with respect to the state $s_i$. Using the symmetry of $W$, the interaction and bias term are derived as follows, equation (17).

$$\frac{\partial}{\partial s_i} \left( -\frac{1}{2}\rho(s)^T W \rho(s) - b^T \rho(s) \right) = -\rho'(s_i) \left\{ \sum_j W_{ij}\, \rho(s_j) + b_i \right\} \tag{17}$$

The potential term is derived as follows, equation (18).

$$\frac{\partial}{\partial s_i} \left( \sum_i \int_0^{s_i} g^{-1}(v)dv \right) = g^{-1}(s_i) \tag{18}$$

Combining those terms yields the partial derivative of the energy with respect to $s_i$ is derived as shown in the equation (19) below.

$$\frac{\partial E}{\partial s_i} = -\rho'(s_i) \left\{ \sum_j W_{ij}\, \rho(s_j) + b_i \right\} + g^{-1}(s_i) \tag{19}$$

Then, the derivation of energy can be written compactly as follows, equation (20).

$$\frac{\partial E}{\partial s_i} = -h_i + g^{-1}(s_i) \tag{20}$$

Let, the evolution of energy along the system trajectory, the time derivative of the energy is calculated as follows, equation (21).

$$\frac{dE}{dt} = \sum_i \frac{\partial E}{\partial s_i} \frac{ds_i}{dt}. \tag{21}$$

Since $s_i = g(u_i)$, we have the equation (22).

$$\frac{ds_i}{dt} = g'(u_i)\frac{du_i}{dt} \tag{22}$$

Substituting the relation and the expression for $\frac{\partial E}{\partial s_i}$ gives the following equation (23).

$$\frac{dE}{dt} = \sum_i \left[ -\rho'(s_i) \left\{ \sum_j W_{ij}\, \rho(s_j) + b_i \right\} + g^{-1}(s) \right] g'(u_i)\frac{du_i}{dt} \tag{23}$$

Because $s_i = g(u_i)$, it follows the following equation (24).

$$u_i = g^{-1}(s_i) \tag{24}$$

Using the dynamics equation (14)

$$u_i + \tau\frac{du_i}{dt} = \rho'(s_i)\left\{ \sum_j W_{ij}\, \rho(s_j) + b_i \right\} \tag{14}$$

, we obtain the time derivative of the energy by following equation (25).

$$\frac{dE}{dt} = \sum_i -\tau g'(u_i)\left(\frac{du_i}{dt}\right)^2 \tag{25}$$

Because the $\rho$ and $g$ are differentiable and monotonically increasing, the time derivative of energy is less than or equal to zero.

$$\frac{dE}{dt} \leq 0 \tag{26}$$

Therefore, the energy function $E(s)$ monotonically decreases along the system trajectories.

This result shows that the proposed nonlinear synaptic Hopfield network admits a valid energy function. The system dynamics always evolve in a direction that decreases the energy, and therefore converge to a stable fixed point corresponding to a local minimum of the energy landscape. Importantly, the nonlinearity introduced at the synaptic level through $\rho$ does not destroy the convergence property of the Hopfield network. Instead, it reshapes the energy landscape while preserving the fundamental Hopfield energy structure. This provides a theoretical justification for using hardware-inspired nonlinear synaptic functions in associative memory systems, as the resulting dynamics remain stable and energy-minimizing.

The expression for the energy gradient also clarifies the role of the regression-based learning rule introduced in Supplementary Note 1. From the gradient expression equation (20), it follows that the stationary points of the energy satisfy the following equation (27).

$$h_i = g^{-1}(s_i) \tag{27}$$

Since $s_i = g(u_i)$, this condition is equivalent to the following equation (28).

$$u_i = \rho'(s_i)\left\{\sum_j W_{ij}\,\rho(s_j) + b_i\right\} \tag{28}$$

Thus, the stationary points of the energy are the fixed-point conditions of the HHN dynamics, where the fixed-point condition follows from equation (1) as described in Supplementary Note 1.

The regression-based learning rule in Supplementary Note 1 is designed to enforce this condition approximately for each stored pattern. Specifically, the learned $W$ and $b$ are chosen such that each stored pattern $t^u$ approximately satisfies the fixed-point relation in its linearized form. As a result, the residual $u_i(t^u) - h_i(t^u)$ becomes small for the stored pattern, and therefore the gradient of energy as described by the following equation (29) also becomes small.

$$\nabla E = \left.\frac{\partial E}{\partial s}\right|_{s=t^u} \tag{29}$$

In this sense, the stored patterns are embedded near stationary points of the HHN energy landscape. Combined with the monotonic energy decrease, this explains why the learned patterns act as stable retrieval states in practice.

**Supplementary Note 3**

Feature-space Interpretation of Nonlinear Synaptic Hopfield Network

In this section, we present a feature-space interpretation of the proposed nonlinear synaptic Hopfield network and clarify its relationship to kernel methods and reservoir computing. This perspective provides insight into how the nonlinear synaptic transformation $\rho$ reshapes the geometry of the associative-memory landscape, thereby enhancing storage capacity.

In the conventional Hopfield network, interactions between neuron states are defined through linear pairwise couplings. The corresponding energy function contains a quadratic term of the form given by the following equation (30).

$$E_{CHN}(s) = -\frac{1}{2}s^T W s \tag{30}$$

The equation (30) implies that the effective interaction between two network states $s_1$ and $s_2$ is governed by the standard inner product $s_1 \cdot s_2$. Under this formulation, the geometry of the state space is fixed by the linear structure of the original neuron coordinates. As a result, the similarity between patterns is measured directly in the original state space, and the representational power of the model is constrained by linear interactions among neuron states.

In contrast, in the proposed HHN, synaptic interactions are not defined directly on the neuronal state vector s, but on a transformed representation obtained through the nonlinear synaptic $\rho$.

The interaction term of the energy function is given by the following equation (31).

$$E_{HHN}(s) = -\frac{1}{2}\rho(s)^T W \rho(s) \tag{31}$$

Therefore, the HHN can be interpreted as performing associative interactions in the transformed feature space defined by $\rho$, rather than directly in the original state space.

Under this view, the effective similarity between two states $s_1$ and $s_2$ is no longer given by a linear inner product, but instead by the following expression (Eq. (32)).

$$K(s_1, s_2) = \rho(s_1)^T \rho(s_2) \tag{32}$$

Thus, the nonlinear synaptic function $\rho$ redefines the similarity metric between network states. The main effect of this transformation is not an increase in the amount of data, but a reconfiguration of the geometry of the state space through a nonlinear change in pattern similarity.

The key role of $\rho$ is to reshape the effective geometry of the state space. Because the similarity between patterns is measured after a nonlinear transformation, the relative distances and overlaps between stored patterns are altered.

This nonlinear geometric reshaping can increase the effective separability of stored patterns in the transformed space. From the perspective of associative memory, this means that attractor basins may become sharper and better separated, reducing overlap between stored memories and suppressing mutual interference.

The improvement arises because the nonlinear transformation changes the geometry of the representation space itself. In this sense, the capacity enhancement of HHN can be understood as a consequence of nonlinear similarity reshaping. This feature-space interpretation is also conceptually related to reservoir computing. In reservoir computing, a nonlinear dynamical system transforms the input into a richer internal representation, and a linear readout layer is then trained to perform the target computation. HHN shares a similar principle: the nonlinear synaptic transformation enriches the effective state representation, while the weight matrix $W$ performs linear associative interaction in the transformed space.

There is, however, an important distinction. In reservoir computing, the nonlinear reservoir and the trained readout are typically treated as separate components. In HHN, by contrast, the nonlinear transformation and the associative dynamics are directly integrated into a single recurrent energy-based system. This makes HHN more compatible with static crossbar hardware, because the nonlinear representation and associative computation are not separated into distinct modules.

From this perspective, the improved storage capacity of HHN can be interpreted as a consequence of nonlinear feature-space reshaping. By redefining pattern similarity through $\rho$, HHN modifies the interaction geometry of the Hopfield energy landscape. This can sharpen the attractor basins corresponding to stored memories and reduce overlap among them, thereby suppressing crosstalk and spurious interference.

This interpretation is consistent with the empirical observation that HHN achieves higher storage capacity than conventional linear Hopfield networks. It should be emphasized, however, that this feature-space argument provides an intuitive and conceptual explanation, rather than a formal proof of capacity bounds. A rigorous theoretical derivation of the exact storage-capacity scaling remains an important direction for future work.

**Uncategorized References**

1 James, G., Witten, D., Hastie, T., Tibshirani, R. & Taylor, J., Linear regression, *Journal*, 69-134 (2023)